# Period Changes and Evolution in Pulsating Variable Stars

**Hilding R. Neilson**
**John R. Percy**
*Department of Astronomy and Astrophysics, University of Toronto, Toronto, ON M5S 3H4, Canada; neilson@astro.utoronto.ca, john.percy@utoronto.ca*

**Horace A. Smith**
*Department of Physics and Astronomy, Michigan State University, East Lansing, MI 48824-2320; smith@pa.msu.edu*

**Abstract**   We review ways in which observations of the changing periods of pulsating variable stars can be used to detect and directly measure their evolution. We briefly describe the two main techniques of analysis—(O–C) analysis and wavelet analysis—and results for pulsating variable star types which are reasonably periodic: type I and II Cepheids, RR Lyrae stars, β Cephei stars, and Mira stars. We comment briefly on δ Scuti stars and pulsating white dwarfs. For some of these variable star types, observations agree approximately with the predictions of evolutionary models, but there still exist significant areas of disagreement that challenge future models of stellar evolution. There may be a need, for instance, to include processes such as rotation, mass loss, and magnetic fields. There may also be non-evolutionary processes which are contributing to the period changes.

## 1. Introduction

Star lives are measured in billions of years, millions of years for very rare massive stars. It might seem, therefore, that it would be impossible for astronomers to observe and measure changes, due to stellar evolution, during their short lives, or during the 400-year life of modern astronomy. But they can! If the star pulsates, then its pulsation period depends on its radius and mass; the period varies approximately as radius$^{1.5}$ and mass$^{-0.5}$, (or period x (mean density)$^{0.5}$ is a constant, called the pulsation constant). Evolutionary changes in one or both of these produce small changes in the pulsation period, which can have a *cumulative* effect on the observed time of the star's maximum (or minimum) brightness.

In this review, we explain how these period changes are measured in several types of pulsating stars, and how they are used, along with theoretical models of the stars, to confirm (or not) our understanding of stellar evolution. At a very basic level, the observations indicate whether the star is expanding or contracting due to evolution, and at what rate. They can also call attention to other processes which may cause period changes, but are not due to evolution. We will highlight ways in which AAVSO observers have contributed, or could contribute, to this area of research.

Note that stars in a more rapid phase of evolution would be rare, but would have larger and easier-to-measure period changes. Stars in a slower phase of evolution would be more common, but would have smaller period changes.

## 2. Methodology

The (O–C) method is the classical method for studying period changes in variable stars, because it is sensitive to the *cumulative* effect of the period changes. It is described in detail in some of the papers in Sterken (2005) and in less detail in the monograph by Percy (2007). It compares the observed time of maximum brightness O with the calculated time C, assuming a known constant period P. When (O–C) is plotted against time, it produces a straight line if the period is constant, a parabola opening upward if the period is increasing linearly, a parabola opening downward if the period is decreasing linearly, and a broken straight line if the period changes abruptly. As an analogy, consider a watch which is running one second more slowly every day. It will lose 1 second in day 1, 2 seconds in day 2, 3 seconds in day 3, 4 seconds on day 4, and so on. The *accumulated* error will be 1 second in day 1, 3 seconds in day 2, 6 seconds in day 3, 10 seconds in day 4, and so on. The accumulated error increases as the *square* of the elapsed time. As noted in section 4.2, there are potential problems if there are long gaps in the dataset. Note also that period changes are expressed in a variety of units, including days per day, days per million years, and seconds per year.

The observed times O can be determined from the observations of the light curve by Pogson's method of bisected chords, or by fitting a standard light curve or an appropriate mathematical function to the portion of the light curve around maximum. If the light curve is stable from cycle to cycle, each cycle of the observed light curve can be fitted to a template made from the average light curve. Since pulsating stars vary periodically in radial velocity, the (O–C) method can be applied to the observed velocity curve also.

The period, and its changes with time, can also be determined by *wavelet analysis*, which is described below.

## 3. Classical Cepheids

3.1. Introduction

The variability of Cepheids was first discovered more than two centuries ago thanks to the naked-eye observations of η Aquilae (Pigott 1785) and of δ Cephei (Goodricke and Bayer 1786). These stars were not the first found to vary, but have ignited centuries of observations of Cepheid variable stars. Along with these stars, Chandler (1893) produced one of the first catalogues of variable stars. That catalogue was soon followed by another (Nijland 1903), which first suggested that the period of δ Cephei was changing with time. Chandler provided a measurement of how much the period was changing in his third catalogue (Chandler 1896, 1904), about –0.05 s yr$^{-1}$.

The discovery that the periods are changing was surprising, but its interpretation was lacking. At the turn of the twentieth century, astronomers argued that Cepheid variability is explained by eclipses in a binary star system (Belopolsky 1895; Campbell 1895). While the period was known to be changing, its meaning was elusive until two more discoveries occurred.

The first of these discoveries was the Leavitt Law (Leavitt and Pickering 1912). The Leavitt Law is the relation between a Cepheid's period and its absolute brightness and is one of the most important tools for measuring distances. Given the distances to the nearest Cepheids, Shapley (1914) showed that Cepheids cannot be eclipsing binary stars, building on the work of Ludendorff (1913), Plummer (1914), and others. If they were eclipsing binary stars, then the radii of the two components would have to be greater than their separation to explain their light curves. The binary hypothesis became absurd, but did persist for another decade (Jeans 1925). Shapley did not offer an alternative hypothesis.

The importance of measuring period change became clear when Eddington (1917) developed the mathematical framework of stellar pulsation. In that derivation, he showed that the pulsation period of a Cepheid depends on the average density within the star. This period-mean density relationship forms the basis for understanding stellar pulsation. Eddington (1918, 1919a) realized that this relationship means that if one measures a change in period then one also measures a change in density. That is, one can measure the evolution of a star from the change in period.

Eddington pointed out the importance of this measurement and implored observers to measure the period change for more Cepheids. He then applied the observations of Chandler, remeasured the period change of the prototype δ Cephei, and tested it against the evolution of stars (Eddington 1919b). Eddington showed that δ Cephei was evolving at too slow a rate for it to generate energy from gravitational contraction. Astronomers had theorized that stars might produce their energy by contraction, but this meant that stars would live for only a few million years. Eddington was able to prove that stars had to generate energy in other ways. This result built the foundation for theories of nuclear energy generation in stars, changing the course of stellar evolution theory.

This one result was crucial for developing ideas of hydrogen fusion in stars, and the groundwork for modern astrophysics. All it required was about one century of watching the variable star δ Cephei. More than 230 years have passed since the discovery of Cepheid variability, allowing for more and more observations. Furthermore, having a longer baseline of observations allows for more precise measurements of period change, hence of stellar evolution.

3.2. Period change in other Cepheids

Period changes have since been measured for other classical Cepheids and, in the next 80 years from 1920 to the new millennium, there are too many studies to cite them all. As such, we highlight some of the most significant developments in measuring Cepheid period change and using the measurements to understand the physics and evolution of these stars.

One of the first leaps ahead was reported by Kukarkin and Florja (1932) for eight Cepheids including the prototype δ Cephei. Even though their sample was small, they presented the first estimate of how rates of period change depend on the pulsation period. They showed that the rate of period change increases as a function of period. This result is striking since two stars in their sample, δ Cep and ζ Gem, have decreasing rates of period change, which means that their densities are increasing. But, it also suggests that the longest-period Cepheids are evolving the most rapidly.

Parenago (1957) measured rates of period change for 24 classical Cepheids plus ten Type II Cepheids. While much of his analysis was consistent with Kukarkin and Florja (1932), he did find evidence for abrupt changes in period that are inconsistent with stellar evolution. There is currently no physical explanation for these abrupt shifts, nor can we be sure that they are real.

While these observations broadened the view of Cepheid period change as function of period itself, it was not until the 1960s when theorists developed the first models of Cepheid evolution and of Cepheid pulsation. These models were another significant leap forward in the understanding of Cepheids (Baker and Kippenhahn 1965; Christy 1963), but they also raised a new set of challenges. For instance, if we can measure how hot and how luminous a Cepheid is, then we can use evolutionary models to measure the star's mass. Similarly, if we measure the period of the Cepheid, we can compute stellar pulsation models to also measure the star's mass. But, it quickly became apparent that stellar evolution models predicted Cepheid masses that were greater than the masses predicted by the pulsation models of the same star. This infamous problem, called the Cepheid mass discrepancy, persists today, though the difference is much less than it was fifty years ago (Cox 1980; Keller 2008; Neilson *et al.* 2011).

In a series of reports, (Hofmeister *et al.* 1964a, 1964b; Hofmeister 1967), Eddington's earlier advice to use Cepheid period change to test stellar evolution was first heeded. Hofmeister realized that having measurements of period change for a large number of Cepheids could be used to test the reliability of the stellar evolution models. This is because, on the Hertzsprung-Russell diagram, stars cross the Cepheid instability strip multiple times and in different directions. For instance, when the pulsation period of a Cepheid is decreasing, the density of the star must be increasing. The density increases when the star shrinks, that is the radius decreases. When the radius decreases the star becomes hotter, hence moves across the Hertzsprung-Russell diagram. In this illustration, the rate of period change tells us about how fast the star is evolving across the Hertzsprung-Russell diagram.

At the same time, the first modern stellar evolution models of Cepheids were produced that suggested Cepheids cross their instability strip at least three times in their lives. This evolution is shown for a star that is five times more massive than the Sun in Figure 1. The star first crosses the instability strip and pulsates relatively soon after the star ceases fusing hydrogen in the core. When this occurs, the star loses a source of energy and the star must expand rapidly to maintain the balance between gravity and pressure. This first crossing is short lasting, from about 100,000 years for the least massive Cepheids and decreasing to about 10,000 years for the most massive. Based on this relatively short time scale, we can expect that Cepheids at this stage of evolution will have rapidly changing periods.

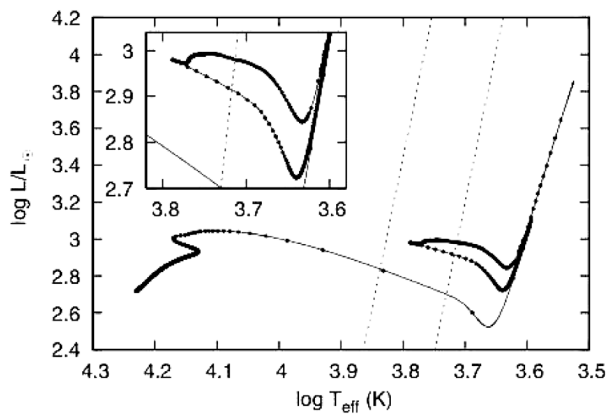

Figure 1. The evolution of a 5 M$_\odot$ star along the Cepheid instability strip on the Hertzsprung-Russell diagram. The dots represent changes in time of 100,000 years. From the number of points, the first crossing of the instability strip lasts less than 100,000 years, while the third crossing lasts about 10 million years. The boundaries of the instability strip were determined by Bono *et al.* (2000).

The star then evolves into a red giant star, reaching its largest size. At the same time the core of the star is contracting and getting hotter. Eventually, nuclear fusion ignites again and the star changes direction in its evolution. The star travels in a blue loop, where the star loops through the Cepheid instability strip leading to the second and third crossings. During the second crossing the star is contracting, hence the mean density is increasing and the pulsation period is decreasing. On the third crossing the opposite occurs. These crossings last from about 10 million years to about 100,000 years for the least to most massive Cepheids, respectively.

These time scales suggest that we can test evolution models if we have measurements of period change for enough Cepheids. Thanks to the work of Hofmeister, a renewed and continuing effort was motivated to measure Cepheid period change. In the thirty to forty years since that work, rates of period change were reported for many more Cepheids, including many instances of measurements for individual Cepheids (for example, Winzer 1973; Erleksova 1978; Fernie 1979). But, of greater interest was the development of programs to gather rates of period change for populations of Cepheids.

One of the first such programs was led by the AAVSO (Cragg 1972). This program gathered observations of about twenty Cepheids as one of the first coordinated efforts to monitor Cepheids. Across the ocean, another program was gathering measurements of period change for about 70 Cepheids with a range of pulsation periods (Szabados 1983). These two analyses offered some of the first measurements of large populations of Cepheids and their period change.

Up to this time, all reports of Cepheid period change were for stars in our galaxy. But, the growth of surveys of Magellanic Cloud stars allowed for astronomers to measure rates of period change for Cepheids in other galaxies. Deasy and Wayman (1985) presented data for classical Cepheids in the Small and Large Magellanic Clouds based on observations spanning from almost 1910 (Payne-Gaposchkin and Gaposchkin 1966; Payne-Gaposchkin 1971) to 1978 (Martin 1981). Period changes were confirmed for about 50 Cepheids in their sample, adding a whole new dimension to understanding Cepheid evolution and pulsation. Because stars in the Magellanic Clouds have less iron, nickel, and elements other than hydrogen and helium in their interiors, their rates of period changes test different conditions for stellar evolution than those of Galactic Cepheids. From their measurements Deasy and Wayman (1985) showed that the evolutionary models developed by Hofmeister (1967) were mostly consistent with the observations. However, Deasy and Wayman (1985) did measure abrupt changes in period beyond the secular changes of period that could not be explained by models.

3.3. The current state of observations

Given the explosion of measurements and surveys since the 1980s, we can now consider the current state of Cepheid period change measurements. This is summarized for two cases: Cepheids in the Galaxy, and Cepheids in the Magellanic Clouds.

Measurements of period change of Cepheids in both the Small and Large Magellanic Clouds have been motivated by large time-series surveys originally designed to search for gravitational microlensing events caused by MACHOs. MACHOs, massive compact halo objects, are small, dim objects that exist in the outer halo of our galaxy. These hypothetical objects would exert gravitational forces on material in the Galactic disk and explain the observed rotation curve of the Galaxy. Hence, detecting MACHOs would solve the dark matter problem (Griest 1991). However, the two main surveys, OGLE (Paczyński *et al.* 1994) and the aptly named MACHO (Alcock *et al.* 1996), failed to detect MACHOs, but they did observe continuously thousands of Cepheids in the Magellanic Clouds. While the MACHO project has ended the OGLE survey is still ongoing, now in its fourth iteration. Poleski (2008) compiled period change measurements from both surveys for both first-overtone and fundamental mode pulsating Cepheids. He found that for observations spanning about 4,000 days, the period change was more pronounced for the first-overtone Cepheids and many of them displayed erratic changes in period that are inconsistent with evolution. These results are surprising and suggest that some other unknown physics is determining the period of Cepheids in these galaxies.

In our galaxy, Turner *et al.* (2006) compiled observations from the Harvard Plate collections and from the AAVSO to measure period changes in almost 200 Galactic Cepheids. These observations spanned about a century and required monumental effort, especially the evaluation of the observation plates. For example, we show a sample O–C diagram for Polaris spanning about 200 years in Figure 2 (Neilson *et al.* 2012a).

These measured rates of period change were found to, again, be broadly consistent with stellar evolution models. While that result might seem underwhelming, Turner *et al.* (2006) presented measurements for a large enough sample of Cepheids as suggested forty years previously by Hofmeister (1967). Neilson *et al.* (2012b) constructed state-of-the-art stellar evolution models of Cepheids and predicted how many Cepheids would have increasing periods and how many would have decreasing periods. This is equivalent to comparing the evolutionary life times for Cepheids evolving on the first and third crossings to those evolving on the second crossing of the instability strip.

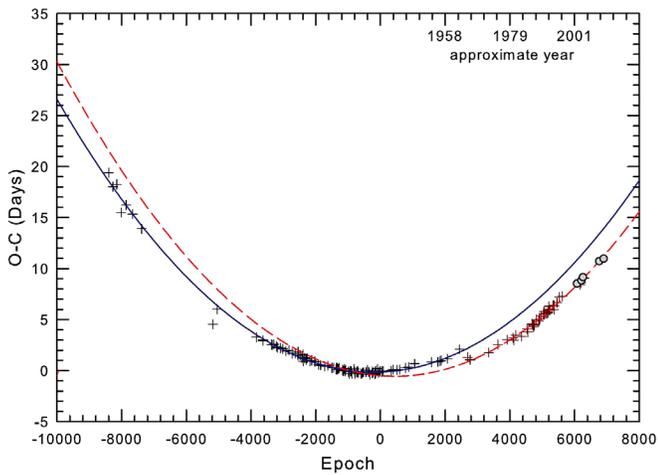

Figure 2. Timing measurements for the Cepheid Polaris over the past 200 years (Neilson *et al.* 2012a). There exists a notable glitch in 1963 that continues to be unexplained. The authors fit the O–C diagram separately over the time frames with lines to note the importance of the glitch. A fit of the period change over all the data yields a measurement of $\dot{P} = 4.47 \pm 1.46$ seconds per year.

These life times are dependent on the physics of the models, such as the nuclear burning rates in the stellar core, rotation and even on their stellar winds. From the Turner *et al.* (2006) sample about two-thirds of Cepheids have positive rates of period change whereas the remaining one-third have negative rates of period change. Assuming so-called standard stellar evolution models, Neilson *et al.* (2012b) predicted that about 85% of Galactic Cepheids should have positive rates of period change. This prediction is much greater than suggested from observations—there appeared to be physics missing from the so-called standard models. The authors computed a new set of stellar evolution models, but this time assuming that during the Cepheid stage of evolution the star underwent enhanced mass loss. This enhanced mass loss has been hinted at by recent infrared and radio observations (Kervella *et al.* 2006; Neilson *et al.* 2010; Marengo *et al.* 2010; Matthews *et al.* 2012). From the new set of evolution models the predicted number of Cepheids with positive rates of period change decreased from 85% to about 70%, much more in line with the fraction suggested by observations. This is the first result to test populations of Galactic Cepheids and their corresponding rates of period change. In one century, Cepheid period changes have been used to disprove gravitational contraction by Eddington to now probing the precise details of their physics. Even more recently, Anderson *et al.* (2014, 2016) computed new stellar evolution models that include the physics of rotation. They showed that rotation impacts the future evolution of Cepheids and that including rotation in the models may be necessary to understand measured rates of period change as well.

While the results from Turner *et al.* (2006) demonstrated the power of period change for understanding stellar evolution, the emergence of new space-based telescopes designed for planet hunting adds yet another dimension. Derekas *et al.* (2012) presented continuous *Kepler* space telescope observations of the one Cepheid in its initial field, V1154 Oph. The time scale of observations was too short to definitively say anything about the evolution of the star but the period was seen to vary by about 30 minutes every pulsation cycle (4.9 days). This variation appeared random and unexplained, based on variations of light of the order of a millimagnitude. Poleski (2008) suggested previously that such variations might be caused by instabilities in the pulsation. Conversely, Neilson and Ignace (2014) showed that the variation might be due to convective granulation in Cepheids. Granules in the Sun are small bubbles of plasma that rise to surface and are both hotter and brighter than the surrounding material. In Cepheids if granules were somewhat bigger (which has been seen in red supergiant stars (Haubois *et al.* 2009)) then the brightness of the Cepheid would vary a small amount from cycle to cycle. Derekas *et al.* (2016) confirmed this hypothesis and quantified the convective granulation properties of the Cepheid. Evans *et al.* (2015) detected similar phenomena in two other Galactic Cepheids from MOST space telescope observations.

The Turner *et al.* (2006) and space-based measurements have allowed for a new path of Cepheid research, but a number of their measurements have since been revisited. Engle *et al.* (2014) and Anderson *et al.* (2015) presented new measurements of the rate of period change for δ Cep that are somewhat different. However, more significant is the revised rate of period change for the long-period Cepheid *l* Carinae. Turner *et al.* (2006) measured the period to be increasing by about 120 seconds every year, but Breitfelder *et al.* (2016) and Neilson *et al.* (2016) independently measured a rate of period change of about 25~seconds every year. Neilson *et al.* (2016) showed that this small rate of period change for such a long-period Cepheid could present significant challenges for understanding this star's evolution. These new results do not diminish the accomplishment of the Turner *et al.* (2006) work, but instead illustrate the need for continued observations of Galactic Cepheids with current precision via the work of the AAVSO.

3.4. The perplexing problem of Polaris

While we could finish the discussion of period change in Cepheids with the work of Turner *et al.* (2006) and the plea for continued monitoring of these stars to improve these measurements, our nearest Cepheid, Polaris, demands its own discussion. Polaris has been a subject of much discussion recently, not just because it is a key step in the cosmic distance ladder, or the celestial navigator, but because it is undergoing changes in its period and brightness that force us to question its evolution. Polaris has been measured to have not only a rapid and positive period change, but its amplitude has also been changing. For much of the past century, the amplitude of light variation in Polaris has been seen to decrease. By the early 1990's Fernie *et al.* (1993) argued that we were seeing Polaris transition from a Cepheid to a "normal" non-pulsating star. However, Polaris did not cease pulsating and its light amplitude appears to be increasing again (Bruntt *et al.* 2008). This variation of amplitude is not at all understood and is one of the open questions about this star.

Another simple and related, yet perplexing question is how far away is Polaris. Van Leeuwen *et al.* (2007) measured a distance of about 130 pc from its HIPPARCOS stellar parallax. However, from its spectrum, Turner *et al.* (2013) measured a smaller distance of about 100 pc. When combined with interferometric measurements of its angular diameter we measure the radius of Polaris. We can take the measured radius as a function of distance and calibrated period-radius relations (Gieren *et al.* 2005) to determine if Polaris is pulsating in the fundamental or first-overtone mode. If Polaris is at the closer distance then it is most likely pulsating in the first-overtone mode. If it is at the farther distance, then Polaris is a fundamental-mode Cepheid.

But, this distance problem is related to our understanding of Polaris from its rate of period change. Turner *et al.* (2005) measured a rate of about 4.5 s yr$^{-1}$ from observations stretching as far back as 1844. This rate is perplexing; Turner *et al.* (2005) noted that this rate of period change challenges the likelihood that Polaris is a fundamental-mode Cepheid because it is too small for a fundamental-mode pulsator on the first crossing of the instability strip yet is too large for a Cepheid on the third crossing. Engle *et al.* (2004) presented independent measurements of the rate of period change of Polaris, yielding similar results to Turner *et al.* (2005). However, the interpretation that Polaris is a first-overtone Cepheid on the first crossing is true only if the HIPPARCOS distance is wrong.

These measurements left a tension between our understanding of the evolution of and distance to Polaris. Neilson *et al.* (2012a) remeasured the rate of period change and again found a similar rate of about 4.5 s yr$^{-1}$. They hypothesized that this tension could be resolved if Polaris were undergoing enhanced mass loss. This enhanced mass loss acts to increase the rate of period change, meaning that Polaris is evolving along the third crossing of the instability strip at the distance measured by van Leeuwen *et al.* (2007). While this idea is interesting it requires Polaris to lose almost one Earth mass of material every year, an amount about 1,000 times greater than understood from standard theory of stellar winds.

This hypothesis, while consistent with recent observations of Cepheid circumstellar environments (Mérand *et al.* 2006; Marengo *et al.* 2010; Matthews *et al.* 2012), is still contentious. As such, we still do not know whether the nearest Cepheid Polaris pulsates as a fundamental or first-overtone mode Cepheid, whether Polaris is about 100 pc away or 130 pc, or whether Polaris is actually undergoing enhanced mass loss in a wind or has a weak stellar wind.

On top of all of these challenges for understanding Polaris, continuous observations show that an unexplained glitch in the pulsation period occurred in 1963 (Turner *et al.* 2005; Neilson *et al.* 2012a). This is seen in the O–C diagram for Polaris, Figure 2 (Neilson *et al.* 2012a). This glitch may be an issue with the timing of observations in 1963 or there may have been some sort of spontaneous change in pulsation period. That type of glitch is still a mystery.

This glitch and the ongoing debate surrounding the period change of Polaris and its properties (Turner *et al.* 2013; van Leeuwen 2013; Neilson 2014; Anderson *et al.* 2016) requires new and continuous observations to help us test theories of pulsation and evolution and to understand the nearest Cepheid.

3.5. Outlook

We are entering an exciting time for probing the connections between pulsation and evolution in classical Cepheids. In the next decade, there will be new space-based telescopes, such as PLATO, TESS, and WFIRST that will take precise observations of Cepheids and measure the details of Cepheid light curves. There will be more continuous surveys from the Earth using facilities such as the Large Synoptic Survey Telescope, along with the continuous observations of OGLE. But, these facilities lack the ability to observe the nearest Cepheids in our galaxy and add to the century or more of time measurements to probe the physics of Cepheid evolution.

There is a continued need for observations of nearby Cepheids to explore the roles of rotation, winds, and other physics. As we learn more about these phenomena then we can also learn more about the evolution of massive stars and supernovae along with helping to precisely calibrate the Cepheid Leavitt Law for measuring the expansion of the Universe.

## 4. RR Lyrae stars and type II Cepheids

RR Lyrae stars and type II Cepheids are both believed to be varieties of old, low-mass stars in advanced but different stages of evolution. We begin with a look at the evolutionary state of RR Lyrae variables. RR Lyrae stars are modest giants, with radii about 4–6 solar radii, pulsation periods in the range 0.2 to 1.0 day, and masses of about 0.6 or 0.7 solar mass. RR Lyrae stars have already evolved off the main sequence, up the red giant branch, and initiated core helium burning at the tip of the red giant branch in a so-called helium flash. After igniting helium burning, they quickly move to the zero-age horizontal branch (ZAHB) where they spend about $10^8$ years deriving energy from the fusion of helium into carbon and oxygen in the stellar core, supplemented by energy from fusing hydrogen to helium in a shell surrounding the helium core. Only those horizontal branch stars that find themselves within the instability strip will pulsate as RR Lyrae variables (Figure 3). A horizontal branch star may thus spend all, none, or only a portion of its core helium burning lifetime as an RR Lyrae. Once they exhaust their central helium, RR Lyrae variables will leave the horizontal branch, eventually exiting the instability strip and becoming red giant stars for a second time, so-called asymptotic branch red giants (Catelan and Smith 2015).

Like other low-mass horizontal branch stars, RR Lyrae stars are very old, older than about $10^{10}$ years. The Milky Way's globular clusters provide examples of such old stellar populations and many, but not all, globular clusters contain RR Lyrae variables. However, more RR Lyrae stars are now known to exist among the field stars of the Milky Way and other Local Group galaxies than are within globular clusters.

Type II Cepheids have periods that range from about 1 day to 25 days. They are sometimes called Population II Cepheids but, while all of the stars within this group appear to be very old, some are metal-rich, unlike true Population II stars. Thus, type II may be a better name. At the upper period limit, the distinction is not always clear between W Vir and RV Tauri variables, and some stars with periods longer than 25 days might legitimately be called type II Cepheids.

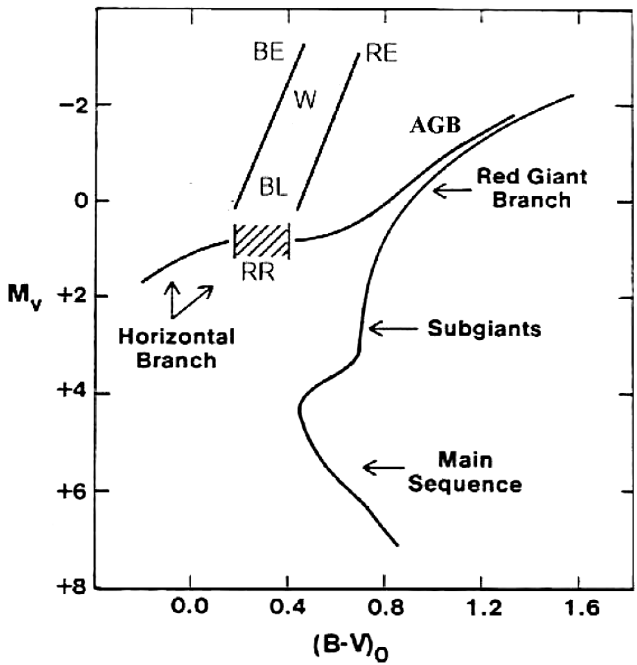

Figure 3. Representative color-magnitude diagram for a globular star cluster. The hatched area on the horizontal branch indicates the location of RR Lyrae variables. BL Her (BL) and W Virginis (W) variables are brighter than RR Lyrae stars and are found between the blue (hot) and red (cool) edges of the instability strip, the slanting lines marked by BE and RE, respectively. The second red giant sequence, labeled AGB, indicates the location of the asymptotic red giant branch.

Like the RR Lyrae stars, type II Cepheids are associated with very old stellar populations, such as globular clusters. They are found within the instability strip, but at luminosities brighter than the level of the horizontal branch (Figure 3). Their masses are smaller and their ages much older than those of the classical Cepheids.

A distinction has long been made between the shorter and longer period type II Cepheids, although different nomenclatures and dividing points have been adopted by different astronomers. Type II Cepheids with periods shorter than 4–8 days have been called BL Herculis stars or AHB1 stars (where the AHB stands for above the horizontal branch), whereas those of longer period are often called W Virginis variables. The *General Catalogue of Variable Stars* (GCVS; Kholopov *et al.* 1985) makes the dividing line 8 days, calling type II Cepheids with periods longer than 8 days CWA variables whereas those of shorter period are CWB stars.

When we consider the evolutionary states of type II Cepheids we find that things are not as well established as for the RR Lyrae variables. A clue to the evolutionary state of type II Cepheids comes from the circumstance that, when they are found within globular clusters, those globular clusters have horizontal branches with a strong component of stars on the blue (hot) side of the RR Lyrae instability strip. This has led to the idea that blue horizontal branch stars are in some fashion the progenitors of type II Cepheids, and that all type II Cepheids have already gone through core helium burning.

For BL Her variables, there is a plausible mechanism by which stars evolving from the blue horizontal branch might enter the instability strip. Blue horizontal branch stars with masses smaller than about 0.51 solar mass are thought to evolve directly to the white dwarf stage. However, theory tells us that blue horizontal branch stars with masses greater than about 0.52 solar mass will become brighter and cooler after the end of their horizontal branch lives, crossing the instability strip from blue to red at luminosities brighter than the level of the RR Lyrae variables. Such stars would occupy the lower portion of the Cepheid instability strip where BL Her variables are located. Eventually they would leave the instability strip to become red giant stars on the asymptotic red giant branch (Gingold 1976; Bono *et al.* 1997; Bono *et al.* 2016).

This evolutionary scenario may explain the occurrence of at least some BL Her variables, but such redward crossing post-horizontal branch stars would not be expected to become as bright as W Virginis variables. An additional crossing of the instability strip at the BL Her level, informally termed "Gingold's nose" after its discoverer, has not been seen in more recent theoretical calculations (Bono *et al.* 2016). In any case, some additional mechanism is necessary to put old stars within the instability strip at the brightness of the W Vir variables.

Unfortunately, the guidance of theory is not entirely clear as to what that mechanism might be. W Virginis variables have sometimes been thought to come from stars that have already reached the asymptotic red giant branch, well to the red of the Cepheid instability strip. It has been proposed that, under the right conditions, asymptotic red giants could undergo short duration loops that move them temporarily blueward into the instability strip while they are undergoing so-called thermal pulses caused by instabilities in their helium burning shells (Schwarzschild and Härm 1970). Thermal pulses are described in slightly more detail in the introduction to section 6. These loop stars would be bright enough to be W Vir stars when they were inside the instability strip. After undergoing such a loop, a W Vir star would, under this scenario, return once more to the asymptotic red giant branch. However, recent calculations have found that most stars with masses appropriate for blue horizontal branch stars may not undergo large thermal instability pulses when they reach the asymptotic red giant stage. Thus, the blueward loop mechanism may not work, or may not work for enough stars to explain the existence of W Vir variables (Bono *et al.* 2016).

Even without thermal pulses, asymptotic red giants would be expected to enter the instability strip at least one more time. At the end of their asymptotic red giant branch lives, they would move from from red to blue on their way to becoming white dwarf stars, crossing the instability strip in the process. This blueward crossing might happen at luminosities brighter than the W Vir level, and the stars might then appear as RV Tauri variables rather than W Vir stars, but it might also be a mechanism for creating W Vir variables. Moreover, blue horizontal branch stars with masses near 0.52 solar mass may not head directly for the asymptotic giant branch after core helium exhaustion, but could possibly briefly loop into the instability strip from the blue side at W Vir luminosities before they reverse course and head for the white dwarf cooling sequence (see Figure 2 in Bono *et al.* 2016).

If that is not confusing enough, there are even more possible sources of complication. The above evolutionary scenarios were modeled for single stars, whereas there is increasing evidence that some type II Cepheids might be members of binary stars (Welch 2012). Differences in photospheric chemical abundances for field type II Cepheids have also suggested that W Vir variables cannot have evolved from stars that were BL Her variables earlier in their existence (Maas *et al.* 2007).

4.1. Theoretical RR Lyrae period changes

The basic pulsation equation guides us in turning theoretical evolution through the Hertzsprung-Russell diagram into changes in period. If the density of a pulsating star decreases, we would expect its period to increase. If the density increases, the period should decrease. A horizontal branch star should slowly change its effective temperature and luminosity as it gradually converts its central helium into carbon and oxygen. However, the lifetime for helium burning by a horizontal branch star is about $10^8$ years, and the resultant changes in density are thus expected to be small over the somewhat more than a century that RR Lyrae stars have been under observation.

Theoretical period changes due to stellar evolution are in almost all cases predicted to be less than about 0.1 day per million years and to occur at nearly constant rates over spans of a century or so (Sweigart and Renzini 1979; Koopmann *et al.* 1994; Kunder *et al.* 2011). Theoretical rates of period change are expected to be largest for RR Lyrae variables at the two extremes of horizontal branch life, the beginning and the end. Toward the end of the life of an RR Lyrae, as it begins to exhaust its central helium, its period might increase at rates of 0.2 or 0.3 day per million years. An RR Lyrae star in those end stages of horizontal branch life might also experience instabilities that could cause temporary period increases or decreases. However, the large majority of RR Lyrae stars would not be expected to be in these final stages of core helium burning.

Larger rates of period change, usually period decreases, might also occur for a small but perhaps non-negligible number of RR Lyrae stars that have not yet begun core helium burning on the horizontal branch but are just about to do so (Silva Aguirre *et al.* 2010). Most such pre-ZAHB RR Lyrae would have period changes between 0 and -1 days per million years, though a few might have even more extreme rates of period change.

4.2. RR Lyrae: observations confront theory

RR Lyrae stars were first established as a class of variable star through the efforts of Harvard astronomer Solon Bailey and his associates, who identified large numbers of them within certain globular clusters (for example, Bailey 1902). However, RR Lyrae itself, still the brightest known member of the class, is a field star that was discovered by another Harvard astronomer, Williamina Fleming (Pickering 1901). Most RR Lyrae stars can be placed into one of two groups: RRab type variables which are believed to be pulsating mainly in the fundamental radial mode, and RRc variables, which are believed to be pulsating in the first overtone radial mode. A smaller number of RR Lyrae variables pulsate in other radial modes or simultaneously in multiple modes (Percy 2007; Catelan and Smith 2015).

By the 1920s and 1930s, it became clear that at least some RR Lyrae variables experienced changes in period (Barnard 1919; Leavitt and Luyten 1924; Martin 1938; Prager 1939). Today, we have observational records for some RR Lyrae stars that extend for more than a century, about 100,000 pulsational cycles. Sometimes period changes of RR Lyrae stars have been discovered through the direct comparison of periods calculated for different years of observation, but more often through an application of some version of the O–C diagram. As noted in the Methodology section above, if the period of a star is increasing or decreasing at a small but constant rate, we expect the O–C diagram to look like a parabola. The rate of period change defined by a parabola can be parameterized by a single number, often called β, which is frequently given in units of days per million years. Note that there is a danger in applying the O–C method to determining the period changes for stars that have significant gaps in the observational record. If an RR Lyrae star is changing rapidly in period, it might not be possible to correctly count the number of cycles that have elapsed between two observed maxima of the star. If the cycle count is wrong, the period change deduced from the O–C diagram will be wrong.

Stellar evolution theory would lead us to expect that the O–C diagram for the large majority of RR Lyrae variables should appear to be either a straight line (no measurable period change) or a parabola corresponding to a value of β smaller than 0.1 day / Myr. A few RR Lyrae near the end of core helium burning might have somewhat larger values of β, up to about +0.3. A few pre-ZAHB RR Lyrae might have large negative values of β. Significant changes in the evolutionary rate of period change would not be expected to occur on a timescale as short as a century.

Observations in many cases contradict this theory. We see too many large and too many variable rates of period change. One example of this is shown by the O–C diagram of the RRab star AR Her. This O–C diagram, shown in Figure 4, is based upon observations obtained between 1926 and 2016 and is calculated assuming a constant period of 0.47000604 day. While the generally increasing O–C until the 1950s and the declining O–C beginning in the 1980s might remind one of rising and falling portions of a parabola, it is clear that something else is going on. Between the 1950s and the 1980s there are abrupt changes in period, both down and up. This diagram further illustrates the danger of gaps in the observational records of RR Lyrae stars. Had no observations of AR Her been obtained between 1960 and 1980 (and the observed maxima are few within that range), we would have no way of knowing how the period of the star fluctuated during those years.

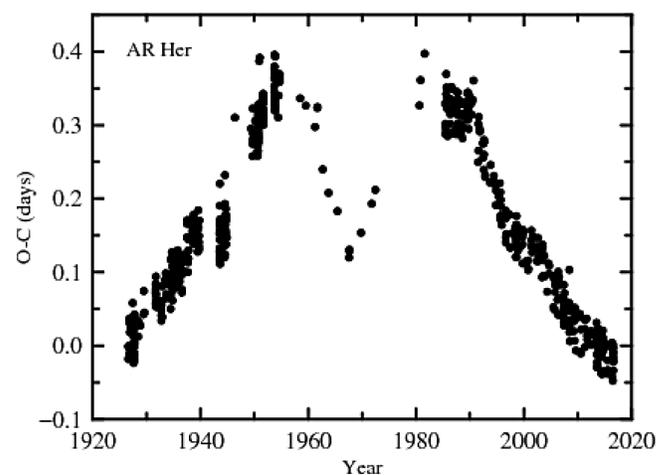

Figure 4. O–C diagram of the RRab star AR Her, courtesy of G. Samolyk and based in part upon AAVSO observations.

O–C diagrams for many field RR Lyrae stars are available on the web through the GEOS project (LeBorgne *et al.* 2007, http://www.ast.obs-mip.fr/users/leborgne/dbRR/). AR Her is not unique. The O–C diagrams of other RR Lyrae stars imply multiple, abrupt period changes, for example, those of XZ Cyg and RW Dra. By contrast, still others have O–C diagrams that imply small or negligible period changes, much more consistent with the predictions of stellar evolution theory. SU Dra and RR Cet would be examples of such variables. In fact, despite the clearly discrepant stars, for many stars the observed rates of period change are consistent with the predictions of stellar evolution theory (LeBorgne 2007; Percy and Tan 2013).

Poretti *et al.* (2016), analyzing 123 RRab stars in the GEOS database, found that 27 stars showed significant period increases, 21 showed signficant decreases, and 75 did not show a significant overall change in period. The median $\beta$ value for the stars of increasing period was +0.14 day/Myr, while the median $\beta$ for the stars of decreasing period was –0.20 day/Myr. A number of RR Lyrae stars had values of $\beta$ more negative than –0.5 day/Myr or greater than +0.5 day/Myr. Most extreme was SV Eri, with a value of $\beta$ near +2.1 (though O–C values for SV Eri show considerable departures from a perfect parabola).

Observed period changes thus suggest that there is some additional source or sources of period change, some sort of period change noise, superposed upon the smooth and generally small rates of period change caused by progressive nuclear burning. We have, then, two rather than one question to address using the observed period changes: (1) do observed rates of period change match those predicted by stellar evolution theory, and (2) what causes the period change noise?

Sweigart and Renzini (1979) suggested that the period change noise might be produced by discrete mixing events in the interiors of the stars that could produce both period increases and decreases. In the long run, these mixing events would average to the evolutionary rate of period change, but that would not necessarily be apparent over a timescale of a century or so. Mass loss on the horizontal branch, if large enough, might also produce additional period changes, but probably not ones as large as frequently observed (Koopmann *et al.* 1994; Catelan 2004). Cox (1998) proposed that period changes might result from the occasional dredging up of helium that had gravitationally settled beneath the convective zones of RR Lyrae variables. Stothers (1980) suggested that the period change noise might have a hydromagnetic origin, associated with changes in the radii of the stars. There are thus plausible mechanisms but no fully agreed upon answer as to which of them actually causes the period change noise.

Must we then give up all thought of testing the stellar evolution of RR Lyrae stars with observed period changes? Perhaps not. One might seek a solution by averaging out random period change noise to reveal the underlying evolutionary changes in period. For any individual RR Lyrae star, one can only be sure of doing this by keeping up observations for many, many years into the future. An alternative, more feasible now, is to average the rates of period change for many RR Lyrae stars. The mean value of $\beta$ for the 123 RRab stars studied by Poretti *et al.* (2016) using the GEOS database is small, as might be expected from slow stellar evolution.

Globular cluster RR Lyrae stars can also help in this endeavor. Some globular clusters contain numerous RR Lyrae stars and have been under observation for a century or more. Perhaps the average behavior of RR Lyrae stars within globular clusters can allow a useful comparison between theory and observation. Again, however, the warning against gaps in the observational record must be given. As we have already noted, gaps can make the correct counting of cycles between points in an O–C diagram difficult, with consequent period change uncertainties.

Individual RR Lyrae variables within globular clusters show a range in period change behavior, as do individual RR Lyrae stars within the field of the Galaxy. However, the average RR Lyrae period changes in different globular clusters is usually close to zero, and in reasonable agreement with theory. Noteworthy is an expected rise in the mean rate of period change as we go to globular clusters with very blue horizontal branches. In these clusters, more RR Lyrae stars than usual would be expected to be evolving from blue to red toward the end of their horizontal branch lifetimes, producing an increase in average $\beta$. However, although this is observed, the higher rate of period increase for blue horizontal branch clusters depends strongly upon the period change results for only a few clusters, especially $\omega$ Centauri (Lee 1991; Catelan and Smith 2015).

As many as half of the RRab stars and perhaps 5–10% of RRc stars exhibit long secondary periods, a phenomenon called the Blazhko effect (Kolenberg 2012). Szeidl *et al.* (2011) and Jurcsik *et al.* (2012) found that RR Lyrae stars in the globular clusters M3 and M5 that exhibited the Blazhko effect were more likely to have erratic period changes. Arellano Ferro *et al.* (2016) note, however, that determining period changes can also be more difficult for such stars. AR Her, with the erratic O–C diagram shown in Figure 4, is also a star with the Blazhko effect. The 123 RRab GEOS stars for which Poretti *et al.* (2016) discussed period changes excluded stars known to show the Blazhko effect, thereby perhaps also excluding some stars with more erratic period behavior.

We note one final use of period change studies of RR Lyrae stars. An RR Lyrae star that is a member of a widely separated binary star can in principle exhibit periodic cycles in the O–C diagram because of differing light travel times when the star is in different portions of its orbit. The number of binary candidates found by this method is still small (Skarka *et al.* 2016), and the possible orbital periods tend to be long (23 years in the case of TU UMa, Wade *et al.* 1999). See also section 5.1 for a further discussion of binary star light-travel time.

4.3. RR Lyrae period changes: outlook for the future

AAVSO monitoring of selected field RR Lyrae variables was begun by Marvin Baldwin (1968) in the 1960s, with observations at first being obtained visually and more recently mainly with CCD cameras. That observing program now continues under the aegis of the short period pulsating variables section of the AAVSO. One of the goals of this program is to provide continued observations of selected RR Lyrae stars, so that there are no gaps in the observational record and it is possible to understand period behavior even for those stars that undergo extreme or erratic variations of period. The AAVSO study of XZ Cygni (Baldwin and Samolyk 2003) shows how even the most extreme period changes can be followed, if there are no gaps in the observational record. Diligent work by AAVSO observers can ensure that the period changes of program stars will continue to be known. A major advantage of the AAVSO community is that, as a group, AAVSO members can target a star for periods of time greater than the observing careers of even the most dedicated individuals.

Long term observations of numerous RR Lyrae stars by programs such as the OGLE survey or the All Sky Automated Survey (Pojmański 1998) will undoubtedly become increasingly important to the study of period changes. An OGLE survey of almost 17,000 RR Lyrae stars in the Galactic bulge found period changes for only 4% of the RRab variables but for 38% of the RRc stars. 75% of the RRc stars with periods in the range 0.35 to 0.45 day showed detectable period changes (Soszyński *et al.* 2011). These results came from observations spanning only some 13 years, and the period change results will undoubtedly become increasingly valuable if the survey is continued for many more years.

4.4. Type II Cepheids: observations confront theory

We saw above that theory predicts that many if not all BL Her stars should show increasing periods, whereas the predicted period changes for W Vir stars are less certain. If blueward loops from the asymptotic red giant branch actually exist, they would generate both period decreases and increases. However, if W Vir stars are mainly evolving blueward in post-asymptotic branch evolution, we would expect period decreases.

Type II Cepheids are a relatively rare type of variable star, and we have available observed period changes for many fewer type II Cepheids than for RR Lyrae stars. In a catalog of variable stars in globular clusters, Clement *et al.* (2001) listed about 2,200 stars with known periods. Of these, about 1,800 were RR Lyrae stars but only about 54 were type II Cepheids (excluding so-called anomalous Cepheids and RV Tauri stars). In Figures 5 and 6 we show observed rates of period change for field and cluster type II Cepheids as a function of period, based upon the determinations in Wehlau and Bohlender (1982), Christianson (1983), Provencal (1986), Holroyd (1989), Wehlau and Froelich (1994), Diethelm (1996), Percy and Hoss (2000), Jurcsik *et al.* (2001), Templeton and Henden (2007), and Rabidoux *et al.* (2010). Period changes for BL Her variables were often determined in these papers by the O–C method, whereas for the W Vir stars a mixture of O–C diagrams and direct determinations of period at different epochs were employed. The timespans covered by the observations run from four decades to a little more than a century, depending upon the star. When more than one paper dealt with the same star, results from the most recent paper with the longest time coverage were adopted. While some researchers provided formal error bars for their period change results, others did not, and the actual errors might depend upon whether cycle counts between observed epochs of observations have been correctly calculated.

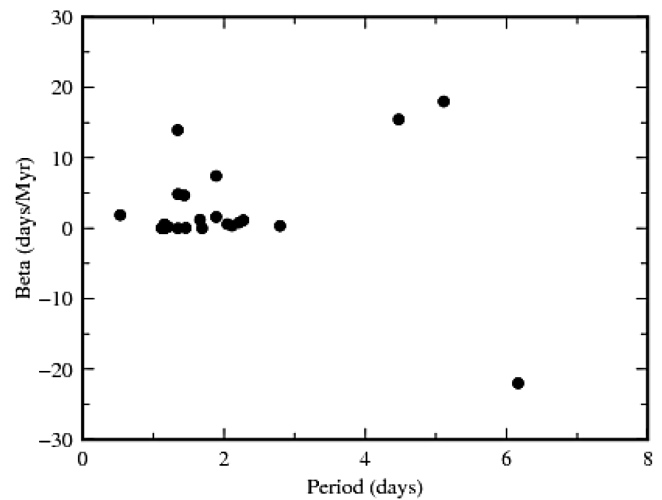

Figure 5. Rates of period change versus period for BL Her stars.

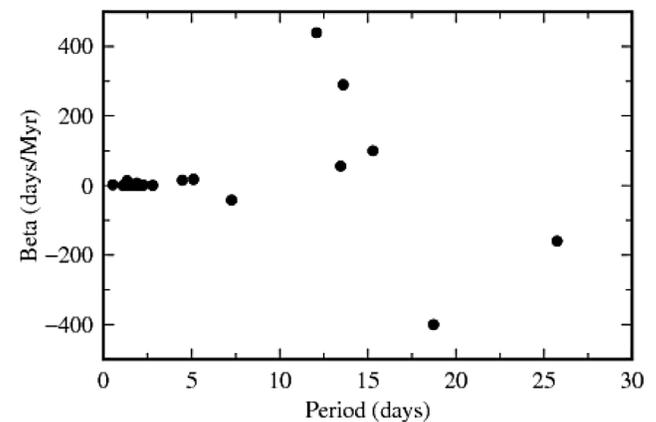

Figure 6. Rates of period change versus period for all type II Cepheids.

In Figure 5 we plot the observed rates of period change for type II Cepheids with periods smaller than 8 days, stars which might fall into the BL Her category. In that figure, almost all of the Cepheids show either small or positive rates of period change. The one star with a significant period decrease is TX Del, which, with a period of 6.165 days, falls at the long period end of the group (Percy and Hoss 2000). When the W Vir stars are added in Figure 6, a dramatically different picture appears. We find very large negative and positive rates of period change.

The observed period changes for type II Cepheids with periods shorter than about 6 days are in line with expectations from stellar evolution theory. Periods for these stars are increasing at rates generally consistent with those expected of post-blue horizontal branch stars evolving to the red. However, it is not clear that the W Vir period changes are consistent with theory. The large rates of period increase and decrease observed for some W Vir stars might be consistent with stars undergoing blue loops from the asymptotic giant branch, but we have seen that there may now be theoretical reasons to question the existence of blueward thermal instability loops. Period increases would not be consistent with the decreasing periods expected in the alternative in which W Vir stars are evolving rapidly to the blue at the end of their asymptotic giant branch lives, but might be consistent with the redward loops into the instability strip predicted for stars with masses near 0.52 solar mass.

We must note, however, that there are complications in determining period changes for W Vir stars. The brighter W Vir stars seem to take on some of the erratic period and phase shift behavior seen among RV Tauri variables (for example, Rabidoux *et al.* 2010). This can introduce extra noise into the O–C diagram. Templeton and Henden (2007) also found that W Vir itself showed pulsation periods in addition to the dominant 17.27-day period. If multiple periods exist for other W Vir stars, that, too, could introduce scatter into an O–C diagram calculated for a constant, single period. As with the other types of variable in this review, there can be complications in interpreting O–C diagrams if there have been large gaps in the observational record.

The complex period behavior possible for W Vir stars can be illustrated by the extreme case of RU Cam (Percy and Hale 1998). In 1965–1966 this seeming type II Cepheid decreased in amplitude from about 1 magnitude to nearly zero. From 1966 until 1982, RU Cam varied with an amplitude of about 0.2 magnitude with a cycle length between 17.6 and 26.6 days, the mean period being about 21.75 days. AAVSO photoelectric observations of RU Cam in 1988–1998 showed a mean period of 22.20 days with variable amplitude. Once more the lesson is that stars sometimes reveal curious behavior when observers assiduously monitor them for long periods of time. The AAVSO should certainly have a role in such monitoring.

## 5. β Cephei stars

5.1. Introduction

The β Cephei stars (also once called β CMa stars), like classical Cepheids, are powerful probes of stellar evolution and structure. These stars, however, tend to be more massive than the classical Cepheids, ranging from about 6 $M_\odot$ to up to $\approx 20_\odot$ (Neilson and Ignace 2015). They are young stars that fuse hydrogen in their cores, similar to the Sun. But, they also exist at the boundary between main sequence blue stars and blue supergiant stars. They have small amplitudes, and require photometric observation.

Because of their evolutionary state, the β Cep stars are powerful laboratories for studying evolution of massive stars. But, even more interesting is the fact that these massive stars are laboratories for more exotic physical phenomena, such as magnetic fields, rotation, and interactions between binary companions. Measurements of period change for these stars allow astronomers to explore this physics.

The β Cep stars have been known to be variable for more than a century. Frost (1906) measured the pulsational velocity variation of the prototype β Cep, and noted the unusually short period. However, it was not for a few decades that period changes were first detected (Struve 1950; Struve *et al.* 1953). Period changes were soon measured for a number of β Cep stars such as δ Cet, 12 Lac, and BW Vul (van Hoof 1965, 1968; Percy 1971).

Using those rates of period change along with other measurements, Eggleton and Percy (1973) conducted the first comparison of theoretical and observed rates of period change. The measured rates of period change ranged from about –1 second every century to about +3.5 seconds every century. Given that the pulsation periods of β Cep stars tend to be about a few hours then the pulsation period will change only a few seconds after almost 300,000 pulsation cycles. Eggleton and Percy (1973) computed stellar evolution models of these stars and predicted their rates of period change to compare to these measured rates. The model rates of period change ranged from about –15 seconds per century to about 100 seconds per century.

While the evolutionary models appeared to be broadly consistent with the measured rates of period change, there was no consensus regarding their state of evolution (Odgers 1965; Percy 1970) nor regarding the physical mechanism that was driving their pulsation. Eggleton and Percy (1973) suggested that β Cep stars pulsated during multiple stages of evolution. Pulsations would begin near the end of core hydrogen burning, a stage where the star is gravitationally contracting, and a third stage where the star burns hydrogen in a shell above the stellar core. The rates of period change reflect the stage of evolution. During the main sequence, rates of period change are small and positive, but during the contraction stage the rates are negative. Finally, during the shell hydrogen-burning stage the rates are greatest.

This analysis appeared to constrain the evolution stages of β Cep stars, but the question of the pulsation-driving mechanism persisted. It had been long-established that pulsations in the classical Cepheid and RR Lyrae stars were driven by ionization of helium in the outer layers of the star. However, Stellingwerf (1978) found that β Cephei pulsation could only be modelled by assuming "enhanced" helium opacities, although the nature of these opacities was unclear. This enhanced opacity mechanism was the leading theory until the study of stellar opacities was revisited (Iglesias *et al.* 1990; Iglesias and Rogers 1991). Moskalik and Dziembowski (1992) found that pulsation is driven by iron ionization in β Cep stars instead of by helium.

The changes in stellar models since the work of Eggleton and Percy (1973) motivated the need to revisit the comparison of theoretical and measured rates of period change. Furthermore, more detailed observations are leading to new measurements of period change. Pigulski and Boratyn (1992) and Pigulski (1992, 1993) revisited the O–C measurements for the prototype β Cep plus σ Sco and BW Vul. In their O–C diagrams, one can fit the standard parabola to measure the secular rate of period change, but when one delves into residual O–C measurements there is a periodic signal (Odell 1984; Jiang 1985). This periodic signal has been argued to be caused by orbital motions of the star in a binary system. In this system the pulsation period of the β Cep star appears longer when the star is moving away from the Earth and shorter when moving towards the Earth since the distance that light travels is changing. This effect is much more commonly observed for pulsar stars in binary systems, but can be seen in a few of these massive stars.

The analysis of Pigulski showed that the O–C and period change measurements can be influenced by binary motions if one is not careful. Along with that analysis, Jerzykiewicz (1999) presented rates of period change that were different from those presented by Eggleton and Percy (1973). The primary difference is that the newer measurements were consistent with positive period change. The combination of new physics in the stellar evolution models and more precise observations motivate new calculations.

### 5.2. Current problems with β Cephei period change

Neilson and Ignace (2015) computed a new grid of stellar evolution models representing β Cephei stars that included physics for rotation and mass loss. It is important to include physical processes such as rapid rotation, magnetic fields, and stellar winds because these effects can help drive evolution in massive stars. Some β Cephei stars have been observed to rotate rapidly (Handler *et al.* 2012) and others slowly (Shultz *et al.* 2015). Some β Cephei stars have powerful magnetic fields (Silvester *et al.* 2009) while others have no detectable field (Fossati *et al.* 2015). All of these observations drive the need for better, more refined stellar evolution models for understanding the lives of massive stars and their deaths as supernovae.

The models computed by Neilson and Ignace (2015) are plotted on the Hertzsprung-Russell diagram in Figure 7 for stars with masses M = 7 to 20 $M_\odot$ that cross the β Cephei instability strip mapped by Pamyatnykh (2007). The models show the effect of different initial rotation rates on the evolution of the stars. Rotation acts to mix material from outer layers in the star into the core and can increase the main sequence lifetime.

When the different models were compared to a sample of period change measurements the results were surprising. The comparison can be summarized by three categories. Those β Cephei stars with measured period changes that were smallest and consistent with no detectable period change agreed with all stellar evolution models. That result simply means that for all stellar masses and physics there is a frame of time where the stars all change their radius and mass, hence period, very slowly. The second category is for the few β Cephei stars with the greatest rates of period change. None of the stellar evolution models and predicted rates of period change could match these measurements implying that there is some phenomenon that we are missing in the models or that the observed period change is not solely due to stellar evolution.

The third category includes measured rates of period change for which the stellar evolution models can be tested. For two β Cephei variable stars, δ Ceti and $\xi^1$ CMa, Neilson and Ignace (2015) used the observed rates of period change to measure the properties of the stars. They found that, thanks to the precision of the measured rates of period change, they could in turn measure both stars' masses and radii, helping to better understand these stars. However, that analysis raised other challenges. For instance, $\xi^1$ CMa is known to have a strong magnetic field (Shultz *et al.* 2015) and δ Ceti is suspected to have a strong magnetic field (Morel *et al.* 2006), whereas none of the β Cephei stars with the greatest rates of period change have strong magnetic fields. Why should the stellar evolution models that do not include magnetic field physics be able to match observations of stars that are strongly magnetic, but not the stars without magnetic fields? It is unclear if there is physics missing from all models or whether the magnetic fields in the β Cephei stars even impact their pulsation. Continued observations are necessary to even begin addressing this complex problem. If we can measure rates of period changes for more than the handful of stars in the Neilson and Ignace (2015) sample, we can start to learn about the impact of magnetic fields and rotation on pulsation and variability in these stars over their evolutionary time scales.

### 6. Period changes in pulsating red giants

When stars with masses of up to about 10 solar masses exhaust the hydrogen fuel in their core, they expand and cool and become *red giant stars*. When the helium in their core becomes hot and dense enough, a "helium flash" occurs; the stars contract somewhat, and begin using helium in their core as fuel. As the helium fuel source is exhausted, the star again expands and cools, and becomes an even more extreme *asymptotic giant branch (AGB) star*, with a radius of up to several hundred solar radii. This process of expansion lasts for hundreds of thousands of years.

Models of AGB star evolution (Wood and Zarro 1981; Iben and Renzini 1983; Boothroyd and Sackmann 1988; Vassiliadis and Wood 1993; Fadeyev 2016) indicate that these stars undergo *thermal pulses* that change their radius and temperature, and possibly their mass, on timescales of hundreds to thousands of years. Thermal pulses result from an instability in the thin shell where helium is being used as fuel. They last in total for a few percent of the AGB evolutionary lifetime, so a few percent of AGB stars should be undergoing such changes, and should therefore show more rapid period changes. AGB stars are highly convective, and convection is a complex and poorly-understood process in cool stars, so the models are somewhat uncertain. Furthermore, the largest AGB stars lose mass at a significant rate, and this further complicates their evolution and the comparison between observation and theoretical models.

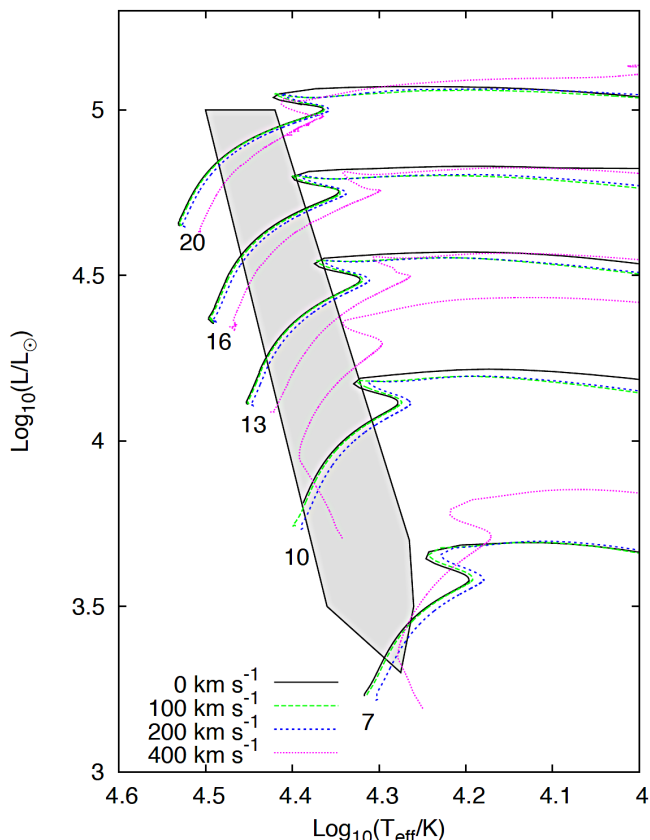

Figure 7. Evolutionary tracks of massive stars on the Hertzsprung-Russell diagram as a function of stellar rotation as the stars pass through the β Cephei instability strip mapped by Pamyatnykh (2007).

In principle, it should be possible to detect the cumulative effect of the slow pulsation period changes in pulsating red giants, due to their evolution up the AGB. This is complicated, first of all, by *random* period fluctuations, described below. Templeton *et al.* (2005) describe some other complications, including the limitations of the visual data on which such studies are usually based. Also, some Mira stars are found in binary systems, and the companion may well affect the structure and evolution of the Mira. Finally, there are other phenomena in pulsating red giants which are not well understood, such as the long secondary periods which are found in about one-third of such stars (Wood 2000), and the variable pulsation amplitudes found in almost all of them (Percy and Abachi 2013). Whatever causes these phenomena may also affect the observed period changes.

Smaller, warmer red giants initially pulsate in a complex mixture of low-amplitude non-radial modes, as the sun does. As the star expands and cools, it becomes unstable to radial pulsation. The period and amplitude are initially a few days and a few millimagnitudes, and the pulsation mode is often an overtone. Later, the period becomes hundreds of days, and the amplitude becomes many magnitudes, and the mode is most often the fundamental. When the visual amplitude exceeds 2.5 magnitudes, the star is by definition a Mira star. Pulsating red giants are classified as M (Mira), SR (semi-regular), and L (irregular). It is very difficult to measure period changes in SR and L stars because of their irregularity and smaller amplitude, but Mira stars, with their large amplitudes, are easy to observe, and reasonably periodic.

AAVSO observers have been systematically observing hundreds of Mira stars for over a century. Kowalsky *et al.* (1986) compiled a database of 75 years of times and magnitudes of maxima and minima of 391 Mira stars, which was used by Percy and others for studies of long-term changes in these stars. More recently, Karlsson (2013) has made public an on-line database on 489 Mira stars, which can be and has been (Karlsson 2014) used for studies of period changes. It includes times of maximum from the literature, and times determined by Karlsson and his colleagues, and other useful information about each star. Ivan Andronov (Kudashkina *et al.* 2014) and his colleagues at the Odessa Observatory, Ukraine, have used additional databases of visual observations to carry out statistical studies of Mira star behavior.

See Smith (2013) for a brief review of period changes in Mira stars up to 2013.

6.1. (O–C) Studies of period changes in pulsating red giants

It has been known for over a century that a few Mira stars show (O–C) diagrams which are parabolic, and which therefore indicate a significant linear change in period. Sterken *et al.* (1999), in elegantly analyzing the change in the period of χ Cyg since its discovery in 1686, have described some of the history of period-change studies of Mira stars. They find a linear increase in the period of χ Cyg, together with quasi-cyclic variations. Wood and Zarro (1981) mention R Aql, W Dra, and R Hya as other early examples of Miras with large period changes.

The (O–C) diagrams of most Miras are not parabolic, but exhibit a meandering or "random walk" appearance (Figure 8). (A random walk is a path which consists of a succession of random steps which may be positive or negative. A simple example is the succession of "heads" and "tails" which result from repeatedly tossing a coin. The steps do not necessarily have to be the same length.) Eddington and Plakidis (1929) showed, for a small number of stars, that this random-walk appearance could be explained by a combination of random, cycle-to-cycle period fluctuations, and random errors in determining the times of maximum. Percy and Colivas (1999) used the Kowalsky *et al.* (1986) database to show that this was true for almost all of the 391 stars, and they determined the size of the average fluctuation for each star. There was a tendency for larger fluctuations to occur in longer-period stars.

The presence of these random fluctuations makes it difficult to observe the slow period increases which should occur due to the slow evolution and expansion of the star. Nevertheless, the slow increases should be present, and might be observed by fitting parabolas to the (O–C) diagrams, determining the rates of period change β, in days per day, and averaging these over a very large number of stars observed over a very long period of time.

Percy and Au (1999) used the Kowalsky *et al.* (1986) database to create (O–C) diagrams for 391 Miras, fit parabolas to these, and average the values of β so obtained. They found that positive period changes outnumbered negative ones; the average value of β was $+16 \times 10^{-6}$ d/d, and the median value was $+14.5 \times 10^{-6}$ d/d. The average value for a model with one solar mass was $+28 \times 10^{-6}$ d/d if the star was pulsating in the fundamental mode and $+11 \times 10^{-6}$ d/d if it was pulsating in the first overtone (Vassiliadis and Wood 1993). The agreement between observation and theory was at about the 1.5σ level. Note that the masses of these stars are quite uncertain, so the predicted rates of period change are known only approximately.

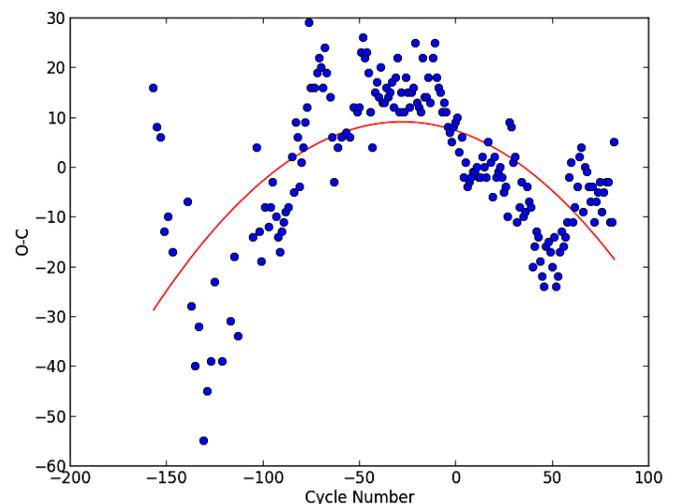

Figure 8. The (O–C) diagram of Z Cap, showing the random-walk pattern caused by the random cycle-to-cycle period fluctuations. The red line is the best-fit parabola, giving a value of β = –0.00455 d/d. Source: Karlsson (2014).

Karlsson (2014) used his database of times of maxima of 489 Mira stars (Karlsson 2013) to study long-term secular changes in period of Mira stars. Karlsson's database is about three decades longer than Kowalsky *et al.*'s. There were 362 stars which had sufficient data for analysis; he constructed (O–C) diagrams for these, and determined values of β. He then applied several tests to these values. (1) The mean period of all the stars increased by 0.15 day over 75 years. (2) Of the stars, 58 percent had increasing periods, and 42 percent had decreasing periods. (3) Almost 60 percent had longer periods in the second half of the dataset than in the first half. (4) The average value of β was $+6.8 \times 10^{-6}$ d/d with a standard error of $3.8 \times 10^{-6}$ d/d. The data are consistent, at about the 2σ level, with the assumption of increasing periods, and with the predictions of models (e.g. Vassiliadis and Wood 1993).

6.2. Wavelet studies of period changes in pulsating red giants

Templeton *et al.* (2005) used a different technique—*wavelet analysis*—to study the period changes of 547 Mira stars in the AAVSO visual observing program. In wavelet analysis, a sine wave of a fixed frequency is fitted to the data using a Gaussian wavelet window function (Foster 1996). In this way, wavelet analysis produces a Fourier transform of each segment of the dataset, giving an estimate of the period(s) and amplitude(s) and how they change with time. A wavelet routine is available within the software package vstar (Benn 2013) on the AAVSO website. Unlike the (O–C) method, which is based on times of maximum, wavelet analysis uses all of the data in the light curve.

Templeton *et al.* (2005) found 57 stars which had period changes significant at the 2σ level, 21 at the 3σ level, and 8 at the 6σ level or higher. The larger period changes are almost equally divided between positive and negative values. As the authors note, the period changes in the 2–3σ range may simply be spurious results of the random fluctuations.

For the stars with the highest rate of period change, the period-versus-time graphs were reasonably linear. For those with lower rates, the graphs were clearly affected by the random fluctuations, and for those with no significant period change, the graphs were random walks. The stars with the largest period changes were: T UMi, LX Cyg, RR Aql, Z Tau, W Dra, R Cen, R Hya, and BH Cru; Templeton *et al.* (2005) discuss these stars individually in some detail; see also Zijlstra *et al.* (2002) for a detailed interpretation of R Hya. Gál and Szatmáry (1995a) have independently analyzed and discussed T UMi. T UMi has also decreased significantly in amplitude, LX Cyg has decreased in mean magnitude by about two magnitudes, and R Hya undergoes large, slow, cyclic variations in mean magnitude. These stars with the highest period changes are assumed to be undergoing thermal pulses. Figure 9 shows the time variation of the period of T UMi, determined by wavelet analysis.

It is alternately possible that T UMi is undergoing a pulsation mode switch from fundamental to first overtone. It is known that some red giants pulsate in the fundamental mode, some in the first overtone, and some in both, so occasional mode switching would not be unexpected. Gál and Szatmáry (1995b) reported possible mode switching in RY Dra, TX Dra, and AF Cyg.

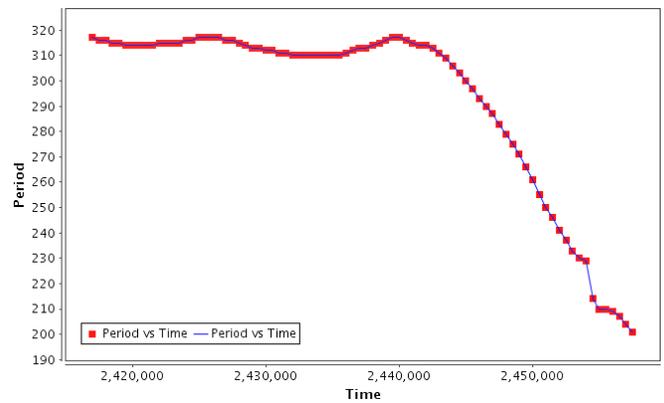

Figure 9. The period of T UMi versus time, determined by wavelet analysis using visual observations from the AAVSO International Database, and the AAVSO software package vstar (Benn 2013). T UMi has the largest rate of period change of any Mira studied by Templeton *et al.* (2005).

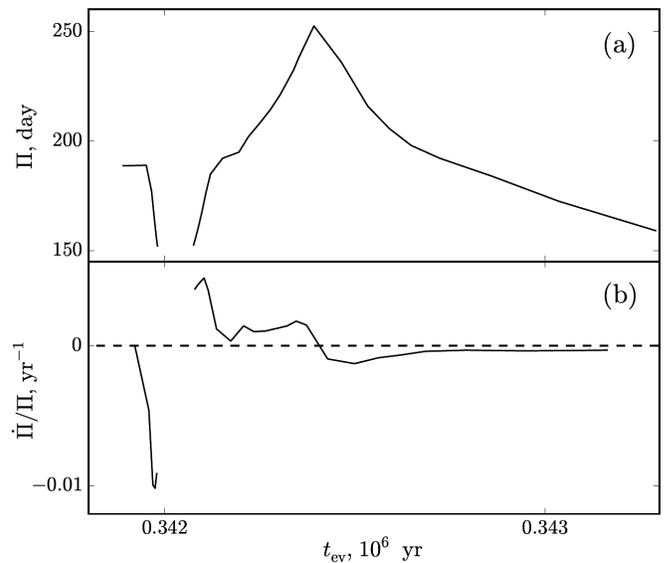

Figure 10. Time dependence of the period in days, and the rate of period change dP/dt/P in years$^{-1}$ for a pulsating red giant model with an initial mass of three solar masses. Significant period changes occur for about a thousand years. From Fadeyev (2016).

Figure 10 shows the time dependence of the period, and rate of period change for a model undergoing a thermal pulse in a red giant with an initial mass of three solar masses (Fadeyev 2016) at a time when the "average" period is about 175 days. This period is somewhat lower than those of most of the stars found by Templeton *et al.* (2005) to have large period changes.

Sabin and Zijlstra (2006) carried out a study, similar to that of Templeton *et al.* (2005), but concentrating on stars with periods longer than 450 days.

Wood and Zarro (1981) call attention to stars, such as T Cep, with small, apparently "abrupt" period changes. Their Figure 4 suggests, however, that the change may simply be part of a random walk.

6.3. V725 Sgr: a unique case

Swope and Shapley (1937) called attention to this remarkable Population II Cepheid which, between 1926 and 1935, increased in period from 12 to 21 days. Percy *et al.* (2006) showed that the star had subsequently and apparently–smoothly changed into a red semi-regular variable with a period of about 90 days. They pointed out that the star's behavior was consistent with a thermal pulse and blue loop in the H-R diagram, from the AGB to the Cepheid instability strip and back again.

6.4. Other observable effects of evolution in pulsating red giants

Since thermal pulses in AGB stars dredge up unstable technetium (Tc) isotopes, several studies have sought to relate the presence of Tc lines in the spectrum to the assumed evolutionary phase and presence of thermal pulses in the stars—Zijlstra *et al.* (2004), for example, who found no obvious correlation between large period changes and spectral features, and Uttenthaler *et al.* (2011), who found a small fraction of the Miras with rapid period changes to show Tc lines; see also the brief review by Whitelock (1999).

Thermal pulses also produce changes in the mean magnitude of the star, and these may be present in some of the hundreds of Miras in the AAVSO program. Since the changes are small and slow, it is essential, for detecting and interpreting them, that the observing procedure, including comparison star magnitudes, stays the same over many decades. Changes in mean magnitude could, however, be due to circumstellar dust formation, rather than to evolution.

Fadeyev (2016) has pointed out that there may also be changes in pulsation *amplitude*, due to the thermal pulses. Percy and Abachi (2013) have shown that semi-regular variables have large, systematic changes in amplitude on a time scale of a few tens of pulsation periods; Mira stars show a similar but smaller effect. These occur on timescales much shorter than the thermal pulse timescale, or the evolutionary one, so they presumably have some other cause. Several other studies have listed Miras which show long-term changes in period, mean magnitude, and/or amplitude—Lebzelter and Andronache (2011), for example, who listed 23 Miras which were candidates for pulsation period (or mode) change, and Percy *et al.* (1990) who used the Kowalsky *et al.* (1986) database, and listed 31 stars with possible changes.

6.5. Prospects for the future

• Red giant stars are highly convective, and convection in cool stars is complex and poorly understood, so evolutionary models could be improved by incorporating an improved theory of convection. This has been done for pulsation models (e.g. Xiong and Deng (2007)) but not yet for evolutionary models.

• The cause of the random cycle-to-cycle period fluctuations is not known, but could and should be investigated using either theory or observation or both. For instance: does the size of the fluctuation correlate with any physical property of the star? Could it be attributed to the effects of large convection cells? Does it reflect some process which occurs on time scales of cycles, or tens of cycles, such as thermal relaxation oscillations in the envelope (Templeton *et al.* 2005)?

• Observational measurements of the slow evolutionary period changes improve, in accuracy, as the square of the length of the dataset. Therefore it is desirable to sustain the systematic observations of as many Miras as possible, since the effects of the random fluctuations need to be averaged out. The present rate of observation seems about right; there is no need to increase the rate for any select group of stars. A perennial question is whether visual observations could be replaced by large-scale surveys with CCD cameras. The times of maximum brightness, determined by visual and CCD V observations, may not be the same, since the brightness of red stars is very sensitive to the wavelength band being used. Any such difference would have to be carefully taken into account.

• Given the present number of stars which have been observed, and the length of the dataset, it may be possible to look for differences in period-change rate and random period fluctuation between stars in different period ranges, and between M and C type red giant stars.

• One of the advantages of systematic observation of large numbers of variables by AAVSO observers is that they often discover stars with unusual behavior, such as significant changes in period, mean magnitude, amplitude, or extreme change as in V725 Sgr. An open question is whether there are any Mira stars which actually undergo *abrupt* period changes. If so, these would not be easy to explain, theoretically.

• Wood and Zarro (1981) proposed that, for R Aql, R Hya, and W Dra, there were changes in mean magnitude which could be explained by the luminosity changes which occur as a result of a thermal pulse. These may be detectable but, since they occur over several decades, it is very important for the observations to be made on the same magnitude system, relative to the same comparison stars, if such a claim is to be valid. This would be one advantage of accumulating sustained, systematic V data on Miras.

• The (O–C) method is a relatively simple one. Papers such as Sterken *et al.* (1999) illustrate the value of using more sophisticated statistical approaches to the problem.

**7. Other types of pulsating stars**

7.1. δ Scuti stars

δ Scuti stars are pulsating stars located in the Cepheid instability strip, on or near the main sequence in the H-R diagram. Most have small amplitudes, and many have two or more radial or non-radial pulsation modes, so they are difficult to observe, analyze, and interpret. A few have larger amplitudes and a single pulsation mode. If they are of Population II, they are usually called SX Phe stars. AAVSO observers frequently observe these larger-amplitude stars, usually using PEP or CCD techniques (see for example, Axelsen 2014).

Their period changes have been reviewed by Breger and Pamyatnykh (1998), and by Templeton (2005) in this *Journal*. They agree that, in most cases, the observed period changes are significantly larger than those predicted by evolutionary models. If the pulsating star is a member of a binary system, some of the apparent period changes may be caused by the light-time effect (Sterken 2005). More often, the large changes are caused by nonlinear interactions between pulsation modes or secular changes in the chemical structure of the stars.

7.2. Degenerate stars

White dwarfs are the inert cores of low-mass stars, exposed at the end of their lives after the stars' outer layers have been cast off as planetary nebulas. White dwarfs have no nuclear energy supply; they shine by slow cooling. As they cool from temperatures of over 100,000 K, they pass through three regimes of pulsational instability, designated DOV, DBV, and DAV. See Kepler *et al.* (2005) for a brief review of white dwarfs and their period changes. Like the δ Scuti stars, they show numerous low-amplitude pulsation modes, so their period changes are also difficult to observe, analyze, and interpret.

In those DOV and DBV stars which have been studied, the observed period changes are greater than models predict, probably due to interaction between the modes. But in a few DAV stars (Kepler *et al.* 2005; Mukadam *et al.* 2013), several decades of high-precision photometry yield period changes in good agreement with evolutionary theory.

## 8. Concluding remarks

We have described how systematic, sustained observations of periodic pulsating variable stars can be used to measure rates of period change in these stars; these, in turn, can be used to detect and measure the slow evolution of the stars. These period changes can be compared with predictions from evolutionary models. Generally, the agreement is good, but there are many cases of disagreement. These disagreements can potentially be used to identify important physical processes, such as rotation, magnetic fields, or mass loss, which need to be incorporated into the evolutionary models.

We have also mentioned how AAVSO observations have helped in the past, and could help in the future. The AAVSO provides a mechanism for making and archiving systematic, sustained observations, which will continue to be useful to astronomers in the future.

## 9. Acknowledgements

We thank the observers who made the observations on which our results were based, including AAVSO observers, and the AAVSO staff who archived the observations. We are grateful to Yuri Fadeyev for permission to use Figure 10. HRN is grateful for instructive discussions with Scott Engle, Edward Guinan, Nancy R. Evans, David Turner, and Richard Ignace. JRP thanks the many students who have worked with him, over the years, on projects related to the subject of this review. HAS thanks Giuseppe Bono for helpful discussions on the theoretical period changes in type II Cepheids.